**Glass transition temperature of PMMA/modified alumina nanocomposite: Molecular dynamic study**

Maryam Mohammadi[a,*], Jamal Davoodi[a], Mahdi Javanbakht[b], Hamidreza Rezaei[d]

[a] Department of Physics, University of Zanjan, Zanjan 45195-313, Iran

[b] Department of Mechanical Engineering, Isfahan University of Technology, 84156-83111, Iran

[d] Department of Materials Engineering, Isfahan University of Technology, Iran

***Corresponding Author**.
E-mail address: mohamady_maryam@znu.ac.ir

## Abstract

In this study, the effect of alumina and modified alumina nanoparticles in a PMMA/alumina nanocomposite was investigated. To attain this goal, the glass transition behavior of poly methyl methacrylate (PMMA), PMMA/alumina and PMMA/hydroxylated alumina nanocomposites were investigated by molecular dynamic simulations (MD). All the MD simulations were performed using the Materials Studio 6.0 software package of Accelrys. To obtain the glass transition temperature, the variation of density vs. temperature was obtained. The temperature at which the slope of the density-temperature curve observably changes is defined as the glass transition temperature ($T_g$). The effect of alumina nanoparticles on the $T_g$ was related to the free volume and the mobility of chain segments and the interaction between the alumina nanoparticles and the polymer. The mobility of the chain segments was investigated based on the mean square displacement and radius gyration. The results show that the increasement the $T_g$ of the PMMA/hydroxylated alumina nanocomposite is more than that of the PMMA/alumina nanocomposite due to the modification of the alumina nanoparticles.

**Highlights:**

- The glass transition of PMMA/modified alumina and PMMA/alumina nanocomposite was calculated
- The mobility, Energy and gyration radius and energy of system was investigated.

# 1 Introduction

Poly methyl methacrylate (PMMA) is a thermoplastic polymer which has been widely utilized in industrial and medical applications due to its advantageous properties such as transparency, impact resistance, resistance to environmental factors, low moisture absorption and biocompatibility [1-3]. To improve the thermal and mechanical properties of the PMMA, fillers such as metal oxides are used [4-7]. Aluminum oxide or alumina is a good candidate for this purpose because of its unique mechanical and thermal properties [8-10]. Many nanocomposite properties are affected by the amount of filler adhesion to the polymer matrix [11-14]. In this study, to increase the adhesion between the fillers and the polymer matrix, hydroxyl alumina and alumina have been used as fillers and their effects on the adhesion were compared.

The glass transition temperature ($T_g$) is an inherent property of polymers at which, a polymer changes from the glass to the rubbery state. Thus, many of its properties change at this temperature [15]. Experimental determination of the $T_g$ requires time and cost. Molecular Dynamics (MD) is a powerful tool by which polymers behavior can be studied at the atomic scale) [16-20]. In this study, the glass transition temperature of the PMMA, PMMA / alumina and PMMA / alumina hydroxyl nanocomposite have been investigated.

# 2 Method

All the simulations were carry out using the Materials Studio6.0 software package of Accelrys based on the molecular dynamics method. In this study, three polymer-based systems including PMMA, PMMA/alumina and PMMA/alumina hydroxyl nanocomposites were studied. In each case, the polymer chain was considered with a degree of polymerization of 50 based on all atom model and the alumina nanoparticle was assumed spherical with a radius of 5 Å (figure 1). Hydroxyl alumina nanoparticles were made by adding hydrogen atoms to the nanoparticle surface. The compass force field was applied for all the atoms in the simulation [21]. The Berendsen thermostat [22] and the Anderson thermostat [23] were utilized to control pressure and temperature, respectively. Periodic boundary conditions were applied to reduce surface effects. The initial density of 0.6 g/cm$^3$ was considered. The MD simulations was followed by allowing the system to equilibrate for 300 ps at a temperature of 1000 K in NPT ensemble. The equilibrium density of the PMMA, PMMA / alumina and PMMA / hydroxyl alumina systems were 1.17 g/cm$^3$, 1.06 g/cm$^3$ and 1.27 g/cm$^3$, respectively.

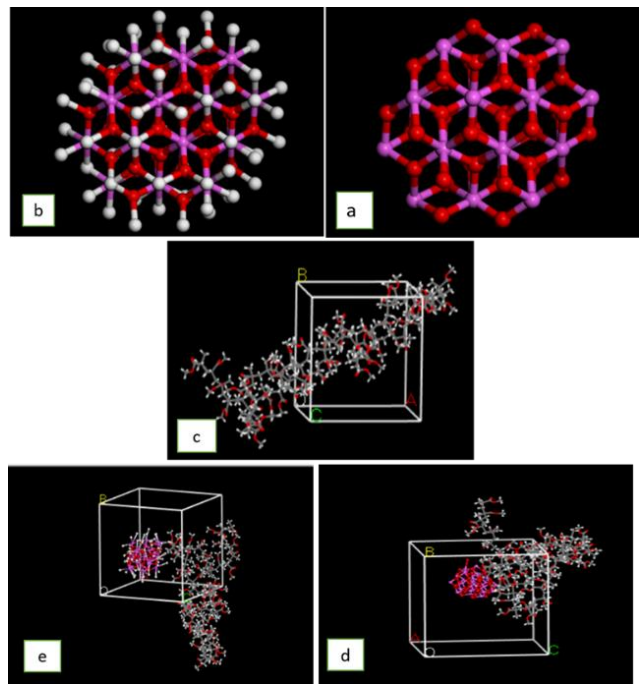

Figure 1. a) Alumina nanoparticle. b) Hydroxylated alumina nanoparticle. c) PMMA chain with 50 monomers. d) PMMA/alumina nanocomposite. e) PMMA/ hydroxylated alumina nanocomposite.

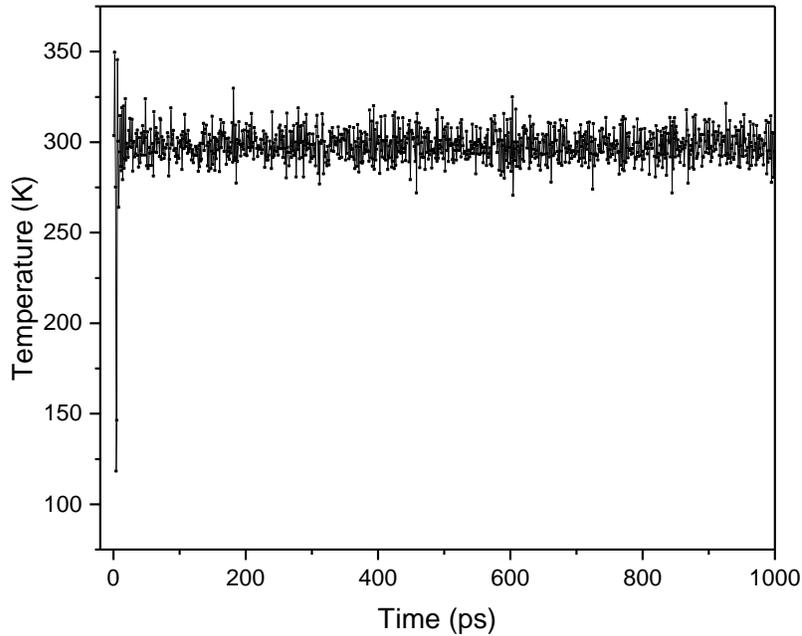

Figure 2. Temperature fluctuations vs. time during the NPT ensemble for the PMMA/alumina.

Next, to remove undesirable interactions annealing was conducted using the NVT ensemble for ten cycles (each cycle 100 ps long). In each cycle, the temperature was increased from 300 K to 1000 K and then was decreased back to 300 K with the temperature step of 20K. The time step of one femtosecond was considered. After annealing the system, the simulation was continued for 500 picoseconds in the NPT ensemble. Figure 2 shows that the temperature fluctuates during time and reaches the constant value of temperature. The energy vs. time is shown in figure 3. As can be seen, the energy fluctuates around a fixed value which represents the equilibrium energy of the system. Next, to determine the $T_g$, the system was cooled from 500 K to 300 K. For any temperature, first, the NPT ensemble with a1000 ps time period and next, the NVT ensemble with a500 ps time period are required to reach the equilibrium.

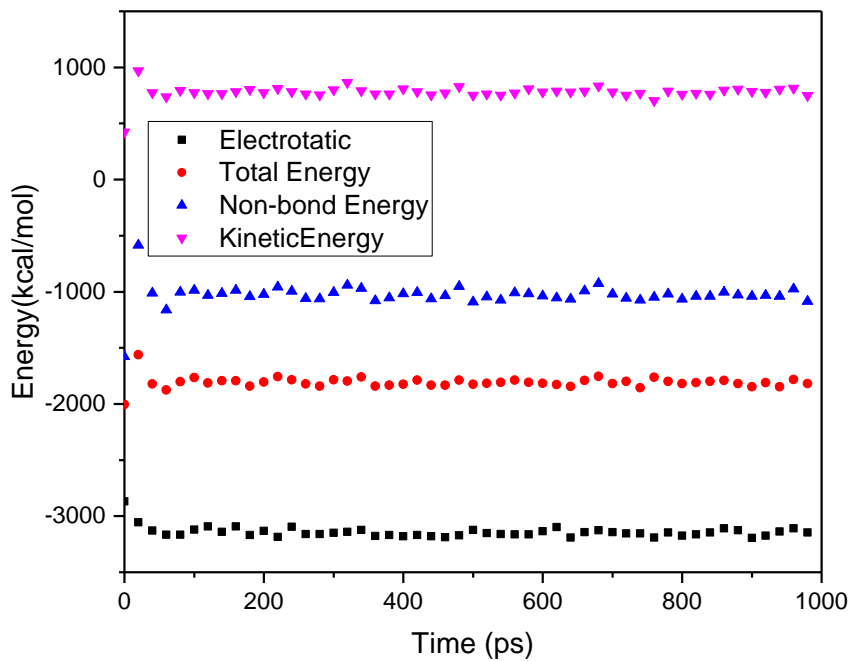

Figure 3. Energy fluctuations vs. time during the NPT ensemble for the PMMA/alumina.

## 3 Results and discussion

### 3.1 The glass transition

Up to now, a complete theory explaining glass transition has not been provided yet [24]. One of the existing theories is the theory of free volume in which the changes in the free volume for the temperatures below the $T_g$ are very small, but for those above the $T_g$ are larger. This is because of the enough energy for molecules in temperatures above the $T_g$. Tg is the temperature at which the change in the free volume reaches a critical value [25]. Thus, a common method for determining the $T_g$ is tracing the density versus temperature. Figure 5 shows the density variations vs. temperature for the PMMA, PMMA/ alumina and PMMA / hydroxyl alumina nanocomposites. For each system, the change in the slop of the density- temperature curve is observable which occurs at the $T_g$. The $T_g$ determined for the PMMA is 450K which is in a good agreement with the MD simulation results[26, 27]. Comparing the three curves shows that the $T_g$ for the PMMA / alumina nanocomposite is $8^o$ K less than that for the PMMA and the $T_g$ for the PMMA / hydroxyl alumina nanocomposite is $12^o$ K higher than that for the PMMA.

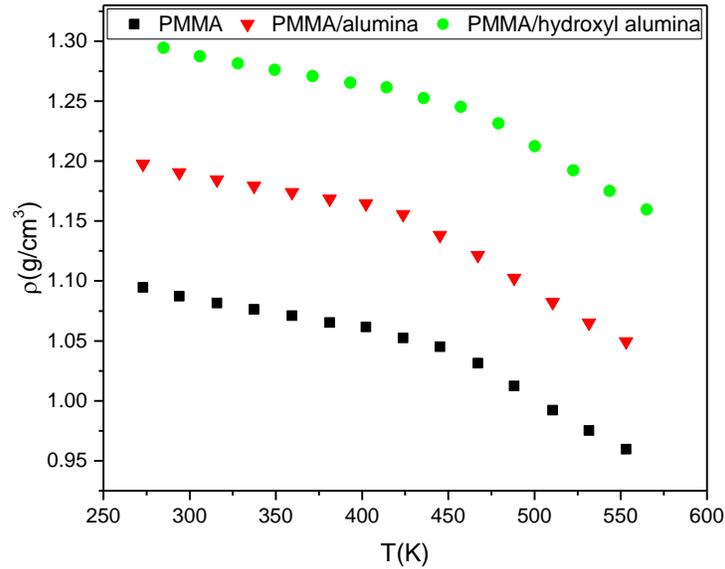

Figure 4. Density vs. temperature for PMMA, PMMA/alumina, PMMA/hydroxylated alumina

In fact, in a nanocomposite, various parameters including the free volume, mobility of the polymer segment and the interaction of the polymer and nanoparticle determine the Tg. Change in temperature in the nanocomposite directly affects the above parameters. Also, a less freedom of motion in the polymer chains is provided the nanocomposite. In the microscopic viewpoint, an increase of alumina particles reduces the free volume and increases the number of cross-linkings between the polymer and the alumina.

### 3.1.1 Chain mobility

The alumina nanoparticle effect on the mobility of the PMMA can be investigated by means of the square displacement and the gyration radius.

**a) Mean-square displacement (MSD)**

The mean-square displacement can be calculated as (equ.1) [28]

$$MSD = \frac{1}{N}\sum_{i=0}^{N-1}(r_i(t) - r_i(0))^2 \qquad (1)$$

where r (0) and r (t) are the positions of atoms at time zero and time t, respectively. N is the number of atoms. As can be seen in Figure 6, the mean square displacement (MSD) for the PMMA / alumina nanocomposite is larger than that for the PMMA. Also, the MSD for the PMMA / hydroxyl alumina nanocomposite is smaller than that for the PMMA.

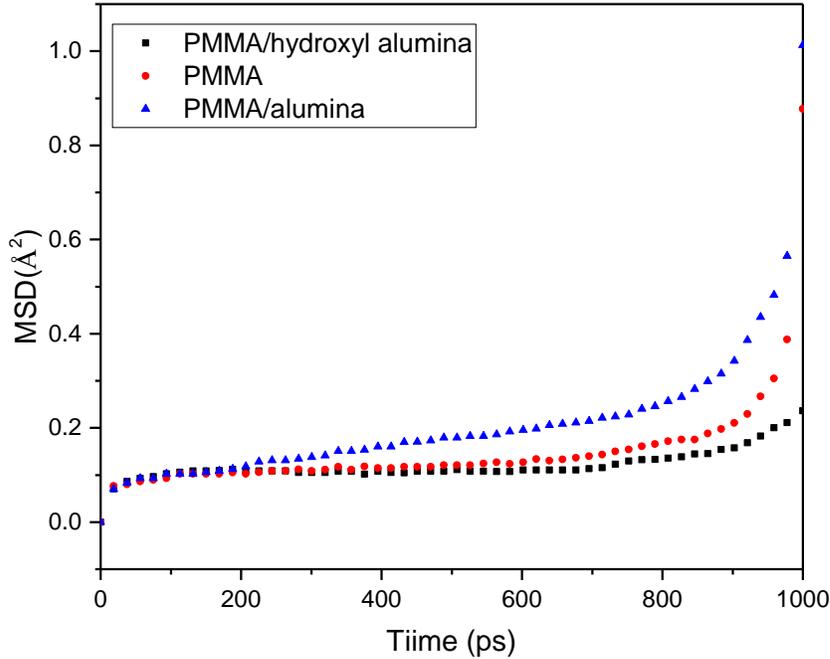

Figure 5. The MSD for the PMMA, PMMA/alumina and PMMA/hydroxylated alumina

**b) Gyration radius ($R_g$)**

The gyration radius shows the shrinkage of polymer chains and can be obtained as [29].

$$R = \sqrt{\frac{\sum m r^2}{\sum m}} \tag{2}$$

where m is the mass of each particle and r is the distance of each atom from the center of mass. The more flexible the polymer chain, the less $R_g$ it has. Calculating the $R_g$ shows that the $R_g$ of the PMMA / alumina nanocomposite is smaller than that of the PMMA. This reveals that the alumina causes change in packaging chains in the intermediate phase. As a result, the change in morphology leads to the difference in thermal properties between pure polymer PMMA and the nanocomposite. The gyration radii for the three systems are shown in Table 1.

Table 1. Gyration radius

| system | Radius gyration(Å) |
|---|---|
| PMMA | 16.5 |
| PMMA/alumina | 15.3 |
| PMMA/hydroxylated alumina | 17.4 |

### 3.1.2 Thermodynamic parameters

The effect of adding alumina nanoparticles on the $T_g$ was related not only to the free volume and mobility of the chain segments, but also to the interaction between alumina nanoparticles and the polymer. Table 2 shows various average thermodynamic properties of the three systems. As it is shown in Table 2, the energy of the PMMA / alumina nanocomposite is smaller than the total energy of the pure polymer. This shows that the addition of alumina particles to the PMMA reduces the interaction between the polymer chains in the PMMA/ alumina nanocomposite. Chain flexibility of the PMMA in the presence of alumina particles is higher than that of the pure polymer and leads to a lower $T_g$. But, for the PMMA / hydroxyl alumina nanocomposite the total energy of the system is larger than that for the pure polymer. This suggests that the interaction between the polymer and alumina is increased by modifying the alumina. The chain flexibility of the PMMA / hydroxyl alumina nanocomposite is lower than that for the pure polymer and increases the $T_g$.

Table 2. Energy of PMMA, PMMA/alumina, PMMA/hydroxylated alumina.

| Energy(kcal/mol) | PMMA | PMMA/alumina | PMMA/hydroxylated alumina |
|---|---|---|---|
| Total | 1764.4 | -1727.3 | -4889.2 |
| Valance(diag. term) | 1442.5 | 1488.1 | 1759.5 |
| Bond | 444.8 | 415.9 | 465.1 |
| Angle | 675.8 | 676.1 | 882.1 |
| Torsion | 213.5 | 290.3 | 323.6 |
| Inversion | 108.3 | 105.7 | 88.7 |

| Energy(kcal/mol) | PMMA | PMMA/alumina | PMMA/hydroxylated alumina |
| --- | --- | --- | --- |
| Valance energy(cross term) | 107.4 | 116.1 | 140.2 |
| Torsion-Stretch | 128.1 | -124.8 | 133.7 |
| Torsion–Bend–Bend | 43.9 | 40.5 | 50.1 |
| Bend–Torsion–Bend | 111.6 | 91.7 | 91.4 |
| Non-bond energy | 3099.4 | 3099.3 | 6508.6 |
| van der Waals | -137.4 | -136.7 | -97.9 |
| Long range correction | -14.6 | -15.7 | -16.1 |
| Electrostatic | 2947.4 | -2946.8 | -6394.7 |

## 4  Conclusion

The effect of alumina and modified alumina nanoparticles on the $T_g$ of the PMMA was studied using the molecular dynamics simulations. To obtain the $T_g$, the variation of the density versus temperature was considered. The temperature at which the slope of the density-temperature curve observably changes was regarded as $T_g$. Also, the addition of the alumina to the PMMA decreases its Tg but adding the modified alumina to the PMMA increases its $T_g$. Investigating the mobility of polymer chains and the thermodynamic properties shows that adding the alumina to the PMMA can increase the mobility of polymer chains, decrease the free volume and reduce the energy system and the interaction between polymer chains. Adding hydroxyl alumina to the polymer reduces the mobility of polymer chains and leads to an increase in the $T_g$. The results show that the MD simulation represented a powerful tool for $T_g$ calculations and can be used for designing and manufacturing new materials.